# Metasurfaces Enable Active-Like Passive Radar

Mingyi Li[1+], Jiawen Xu[1+], Hanting Zhao[1+], Xu Zhao[1], Yan Jin Chen[1], Tie Jun Cui[2,3], Vincenzo Galdi[4], and Lianlin Li[1,5]

[1] State Key Laboratory of Photonics and Communications, Peking University, Beijing 100871, China
[2] State Key Laboratory of Millimeter Waves, Southeast University, Nanjing 210096, China
[3] Suzhou Laboratory, Suzhou 215004, China
[4] Fields & Waves Lab, Department of Engineering, University of Sannio, I-82100 Benevento, Italy
[5] Pazhou Laboratory (Huangpu), Guangzhou, Guangdong 510555, China

## Abstract

Passive radars (PRs) provide a low-cost and energy-efficient approach to object detection by reusing existing wireless transmissions instead of emitting dedicated probing signals. Yet, conventional passive systems require prior knowledge of non-cooperative source waveforms, are vulnerable to strong interference, and rely on Doppler signatures, limiting their ability to detect subtle or slow-moving targets. Here, we introduce a metasurface-enabled PR (MEPR) concept that integrates a space-time-coding programmable metasurface to imprint distinct spatiotemporal tags onto ambient wireless wavefields. This mechanism transforms a PR into an *active-like* sensing platform without the need for source control, enabling interference suppression, signal enhancement, and accurate target localization and tracking in cluttered environments. A proof-of-concept implementation operating at 5.48 GHz confirms real-time imaging and tracking of unmanned aerial vehicles under interference-rich conditions, with performance comparable to active radar systems. These results establish MEPR as a solid foundation for scalable, adaptive, and energy-efficient next-generation integrated sensing and communication systems.

# Introduction

Passive imaging is pervasive across natural phenomena and technological applications, spanning the full electromagnetic (EM) spectrum from optics to radio waves, as well as seismic, ultrasonic, and other wave-based systems [1-3]. It allows observation and reconstruction of the world by exploiting naturally existing illumination rather than transmitting dedicated probing signals. In the radio regime, passive radars (PRs) exploit existing wireless transmissions, such as broadcast television, cellular signals, or WiFi, to detect and track targets [2]. In contrast to conventional active radar systems, PRs offer several advantages, including low power consumption and deployment cost, inherent concealment, and resilience to jamming [3-7]. Such appealing features make them attractive for diverse applications including defense surveillance, urban monitoring, and traffic management, where they can take advantage of existing communication infrastructure to provide cost-effective and scalable sensing. However, the passive nature of these systems presents significant challenges. Detection relies heavily on Doppler signatures, which reduces sensitivity to slowly moving targets. The limited aperture constrains spatial resolution and sensitivity to small objects, while strong direct-path and co-channel interference can overwhelm weak target echoes, thereby reducing detection accuracy. Moreover, because PRs depend on non-cooperative sources, receivers typically require demodulation and structural analysis of the transmitted waveform to obtain a reference, which complicates implementation and restricts practical applicability. Overcoming these difficulties requires approaches that can suppress interference and extract target echoes without prior knowledge of the source signals.

Recent advances in artificial EM materials provide new opportunities to overcome these limitations. Metasurfaces, composed of two-dimensional arrays of subwavelength structures, enable precise control of EM waves [8], supporting effects such as anomalous reflection and refraction [9] and cloaking [10]. When equipped with active components such as positive-intrinsic-negative (PIN) diodes [11,12], varactor diodes [13], or micro-electro-mechanical system (MEMS) switches [14], such structures are commonly referred to as reconfigurable intelligent surfaces (RISs) [15], which bridge the physical and digital domains and extend the study of metasurfaces into the realms of information science [16] and artificial intelligence [17]. Demonstrated applications of these programmable metasurfaces (PMs) include beam steering [12,15], holography [18], and imaging [19]. Space-time-coding PMs (STC-PMs) extend these concepts by enabling dynamic control of EM waves across both spatial and frequency domains [20–24]. This spatiotemporal modulation introduces a new information dimension to wave-matter interactions, allowing signals to be selectively tagged, redirected, or filtered in real time. When incorporated into radar architectures, particularly PR systems, such PMs provide a physical mechanism for echo separation and interference suppression without the need to demodulate the source signal. These advances are particularly relevant to low-altitude airspace security, where unmanned aerial vehicles (UAVs) have become increasingly challenging to monitor. Their small radar cross section, slow motion, and complex flight environments make them elusive to conventional PRs, which struggle to extract weak echoes from clutter and interference [25–28].

In this work, we introduce a metasurface-enabled PR (MEPR) concept that redefines passive sensing by embedding spatiotemporal modulation signatures directly into ambient wavefields. This approach enables the selective tagging and isolation of target echoes without prior knowledge of non-cooperative wireless transmissions, turning ambient wireless signals into structured illumination for sensing. A proof-of-concept prototype demonstrates real-time UAV tracking with sub-meter accuracy under strong interference, establishing metasurface-driven spatiotemporal control as a foundation for a new class of intelligent, adaptive, and energy-efficient radar systems.

## Results

**MEPR principle.** We begin by describing the operating mechanism of a MEPR, which employs an STC-PM to convert ambient wireless signals into structured sensing waveforms. By dynamically modulating the phase of the incident field across space and time, the metasurface embeds unique spatiotemporal tags into the illumination directed toward the target. These tags are imprinted on the backscattered echoes and serve as identifiers that distinguish the desired reflections from strong direct-path and co-channel interference. As shown in **Fig. 1a**, the MEPR is implemented to detect and track a UAV in a representative urban environment. The system consists of an STC-PM mounted on a building facade and two coherent receivers: a reference receiver $R_1$ at position $\boldsymbol{r}_1$ and a surveillance receiver $R_2$ at $r_2$ (see details in the **Methods** section and **Supplementary Note 1**). The signals collected by the two receivers are jointly processed through code-correlated reconstruction, enabling robust target localization and tracking even in interference-rich conditions.

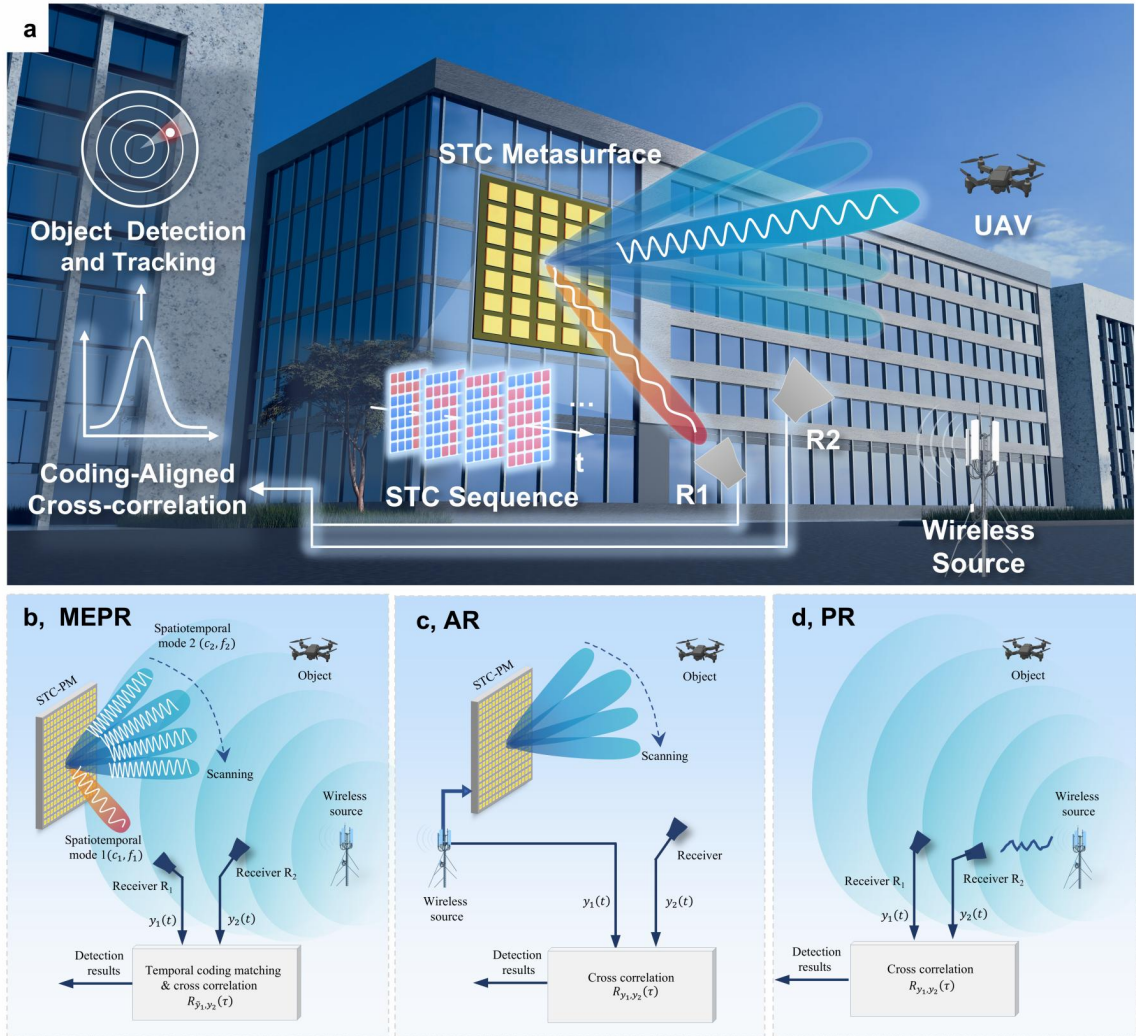

**Figure 1 | Conceptual illustration of an MEPR system.** **a,** Conceptual overview of the MEPR used for detecting and tracking a UAV in a complex urban environment. The system consists of an STC-PM mounted on a building facade, two synchronized receivers (a reference and a surveillance channel), and a post-processing module. The STC-PM dynamically modulates ambient wireless signals in space and time, embedding distinct temporal tags that enable the separation of target echoes from environmental interference. **b,** Schematic of the space-time modulation mechanism. The STC-PM converts incoming wireless signals into two tagged components: one defined by $(c_1, f_1)$ to generate the reference signal $y_1(t)$ at receiver $R_1$, and the other defined by $(c_2, f_2)$ to produce the scanning signal $y_2(t)$ for the region of interest. The post-processing stage performs temporal code matching and cross-correlation between $y_1(t)$ and $y_2(t)$ to extract the target response. **c,** Comparison with a conventional active radar, in which the transmitted waveform is known and coherently linked to the transmitter, allowing direct

detection. **d,** Comparison with a traditional PR lacking STC modulation, where direct-path and co-channel interference remain dominant, hindering the detection of weak or slow-moving targets.

In essence, the STC-PM functions as a physical spatiotemporal modulator that shapes the propagation of ambient wireless signals in both space and time. Its overall response can be expressed as: $\mathcal{H}(t,\boldsymbol{r}): s(t) \to y(t;\boldsymbol{r})$, where $s(t)$ denotes the incident field and $y(t;\boldsymbol{r})$ represents the field observed at position $\boldsymbol{r}$. This relationship can be described by a $K$-component spatiotemporal modal decomposition,

$$y(t;\boldsymbol{r}) = \sum_{k=1}^{K} c_k(t - \tau_{m \to r})[f_k(t - \tau_{m \to r}; \boldsymbol{r}) \otimes s(t - \tau_{m \to r})], \quad (1)$$

where each temporal mode $c_k(t)$ ($0 \leq t \leq T_L$, with unit energy, i.e., $\|c_k(t)\| = 1$) governs the time-varying reflection of the incident signal, and each spatial mode $f_k(t;\boldsymbol{r})$ determines the corresponding radiation pattern. The operator $\otimes$ denotes convolution over time, and $\tau_{m \to r}$ represents the propagation delay from the metasurface to the observation point. The total coding duration $T_L = MT_c$ consists of $M$ temporal segments of length $T_c$. The temporal modes vary slowly compared with the spatial modulation, so that the metasurface effectively "writes" time-varying tags onto different beams. These tags are chosen to be mutually uncorrelated, typically implemented using chirp or random sequences, so that each tagged reflection can be uniquely identified during correlation processing. Through this mechanism, the MEPR converts unstructured ambient radiation into information-carrying wavefronts, enabling passive sensing with the precision of active radars. Further details on Eq. (1) are provided in the **Methods** section and **Supplementary Note 2**.

For clarity, we focus on the beam-scanning operation of an MEPR, which is central to object detection in this study. A narrow and steerable beam is required to scan the region of interest efficiently. To achieve this, the STC-PM is configured to generate two distinct spatial modes, $f_1$ and $f_2$, as illustrated in **Fig. 1b**. The first mode establishes a stable wireless link between a selected ambient source and the reference receiver $R_1$, producing a temporally coded reference signal. The second mode forms a scanning beam directed toward the target region, enabling the detection of reflected echoes. These two modes correspond to the coded components $c_1(t - \tau_{m \to r_1})s_1^h(t - \tau_{m \to r_1}; \boldsymbol{r}_1)$ and $c_2(t -$ $\tau_{m \to r_u})s_2^h(t - \tau_{m \to r_u}; \boldsymbol{r}_u)$, representing the STC-PM-modulated illuminations toward the reference receiver and the target, respectively.

The received signals at the two receivers can be written as

$$y_1(t) \approx G_1 c_1(t - \tau_{m \to r_1})s(t - \tau_{m \to r_1}) + n_1(t), \qquad (2)$$

$$y_2(t) \approx \alpha_u G_2 c_2(t - \tau_{m \to r_u \to r_2})s(t - \tau_{m \to r_u \to r_2}) + n_2(t), \qquad (3)$$

where $n_1(t)$ and $n_2(t)$ include all signal components unrelated to the STC-PM modulation, such as direct-path interference. Here, $\tau_{m \to r_1}$ and $\tau_{m \to r_u \to r_2}$ denote the propagation delays from the metasurface to receiver $R_1$, and from the metasurface to receiver $R_2$ via the target, respectively; $\alpha_u$ is the target reflection response; and $G_1$ and $G_2$ are the effective radiation gains associated with the two spatial modes.

After temporal-code projection, defined as $y_1 \to \tilde{y}_1 = y_1 c_1^* c_2$, with the asterisk denoting complex conjugation, the signal models in Eqs. (2) and (3) become consistent with those of an active radar. In this form, cross-correlation effectively suppresses the unwanted components $n_1 c_1^* c_2$ and $n_2$, allowing an MEPR to extract target echoes with high contrast even in the presence of strong interference. Receivers $R_1$ and $R_2$ are directionally aligned toward the metasurface and the target region, respectively. In addition, receiver $R_2$ is positioned and configured in its antenna gain pattern to minimize sensitivity to the spatiotemporally modulated reflections from the metasurface. Further derivation details of Eqs. (2)-(3) are provided in **Supplementary Note 3**.

The object detection process in the MEPR is carried out by cross-correlating the temporally projected reference signal $y_1 c_1^* c_2$ with the received signal $y_2$, expressed as

$$R_{\tilde{y}_1, y_2}(\tau) = \frac{\left|\int [y_1(t) c_1^*(t - \tau_{m \to r_1}) c_2(t - \tau_{m \to r_1})] y_2^*(t + \tau) dt\right|}{\left|\int [y_1(t) c_1^*(t - \tau_{m \to r_1}) c_2(t - \tau_{m \to r_1})] dt\right|^2}$$

$$\propto \frac{G_2^*}{G_1} \alpha_u^* R_{c_2 s, c_2 s}(\tau - \tau_{m \to r_u \to r_2}), \qquad (4)$$

where $R_{c_2 s, c_2 s}$ is the autocorrelation function of the composite signal $c_2(t)s(t)$. In deriving the proportionality in Eq. (4), it is assumed that the cross-terms involving noise and unmodulated background signals satisfy

$\int |s(t)|^2 dt \gg \left|\int c_2(t) n_2^*(t) s(t) dt\right|, \left|\int c_1^*(t) n_1(t) s^*(t) dt\right|, \left|\int c_1^*(t) c_2(t) n_1(t) n_2^*(t) dt\right|,$

which hold with high probability under mild conditions (see **Supplementary Note 4**). This formulation reveals the central mechanism of the MEPR: the temporal modes $\{c_k(t)\}$ generated by the STC-PM imprint unique modulation signatures onto the scattered fields, enabling the selective extraction of target echoes from strong interference. Equation (4) indicates that, in essence, an MEPR behaves analogously to an active radar system transmitting the coded waveform $c_2(t)s(t)$ through the metasurface aperture. Consequently, the achievable azimuth resolution is mainly governed by the effective aperture size of the STC-PM, while the range resolution is determined by the signal bandwidth of $c_2(t)s(t)$. Since ambient wireless signals $s(t)$ typically have limited bandwidth, the range resolution is inherently coarser than that of conventional wideband active radars.

Before concluding this section, we discuss how the MEPR concept relates to conventional active and passive radar systems, and highlight its distinct advantage. From a hardware perspective, the MEPR becomes equivalent to an active radar when the ambient signal $s(t)$ is fully cooperative with the receiver and directly connected to the transmitting antenna of the STC-PM, as illustrated in **Fig. 1c**. Under this condition, and for sufficiently long temporal coding duration $T_L$, all terms unrelated to the STC-PM modulation, $n_{1,2}(t)$, can be efficiently suppressed. The resulting signal model and detection principle are thus nearly identical to those of an active radar, which explains why the MEPR operates in an "active-like" manner despite using ambient signals.

Conversely, if the STC-PM is removed, the MEPR reduces to a conventional PR, as shown in **Fig. 1d**. In this case, the system faces two intrinsic limitations. First, accurate knowledge of the source waveform is required to generate a coherent reference channel, which contradicts the core idea of passive operation. Second, the weak echoes from small or slow-moving targets are often overwhelmed by strong direct-path and co-channel interference, severely degrading detection performance in complex EM environments. In contrast, the STC-PM in the MEPR introduces controllable spatiotemporal modulation that embeds identifiable signatures into the reflected field. This capability enables the system to isolate target echoes from ambient interference, effectively overcoming the limitations of conventional PRs while retaining their low-power and source-independent nature.

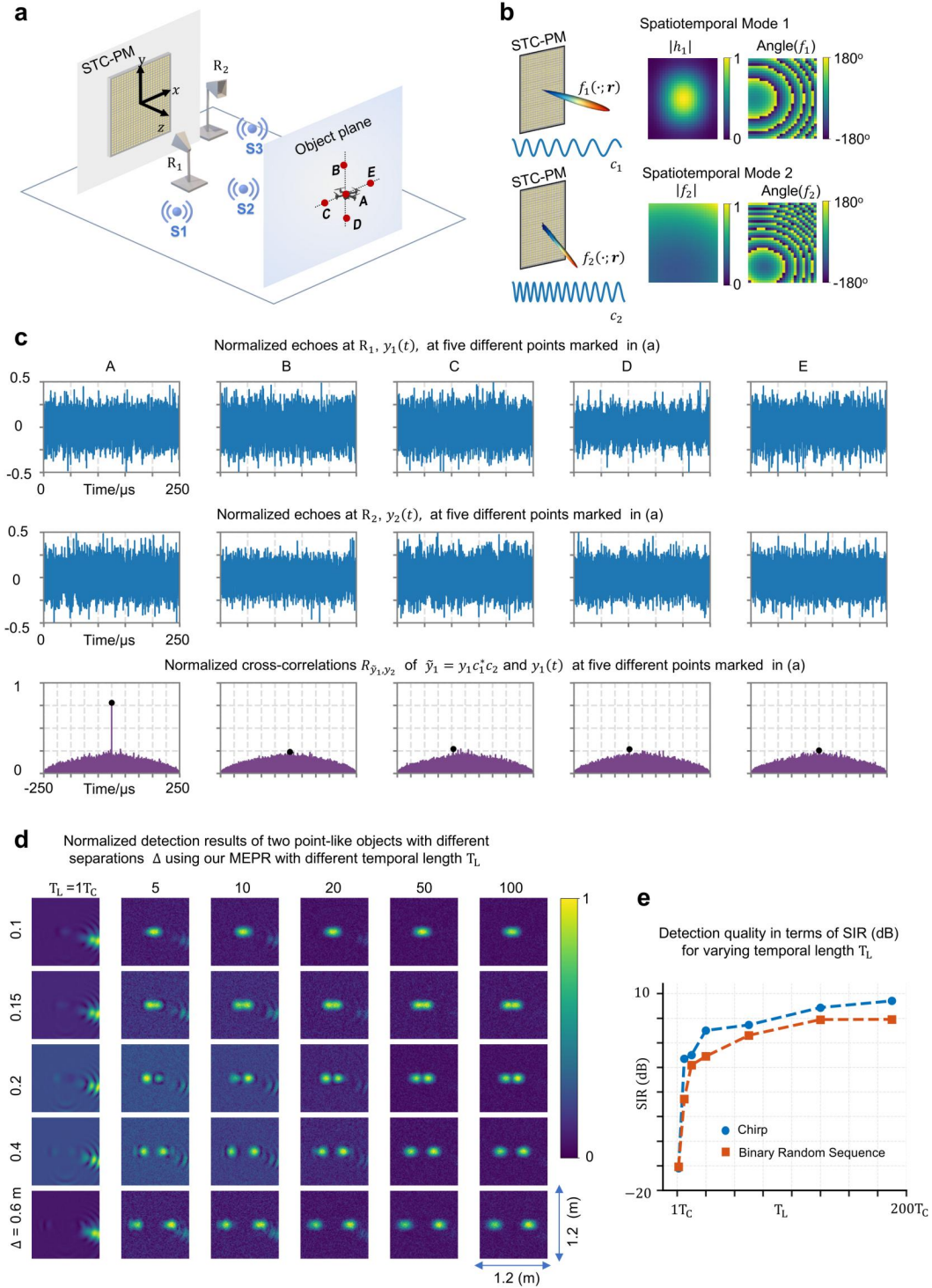

**Figure 2 | Numerical verification of the MEPR principle.** Simulations are performed for an STC-PM of size 0.781 × 0.586 m², consisting of a 32×24 array of meta-atoms, each with continuously tunable amplitude and phase responses in the ranges [0, 1] and [0°, 180°], respectively. **a,** Numerical model of the MEPR configuration, where the STC-PM is placed at the origin and two receivers are located at

designated positions. Multiple non-cooperative wireless sources are randomly distributed around the scene to emulate a realistic interference-rich environment. **b,** Representative spatiotemporal modes generated by the STC-PM: one mode directs a reference beam toward the reference receiver, while the other performs dynamic beam scanning across the object space. The corresponding coding matrices are shown in **Supplementary Video 1**. **c,** Simulated echo signals collected by receivers $R_1$ and $R_2$ and their cross-correlation functions $R_{\tilde{y}_1,y_2}(\tau)$, corresponding to five probe points (A-E) marked in **a**. A point-like object located at A produces a distinct correlation peak when the scanning beam $(c_2, h_2)$ is directed toward it. **d,** Simulated detection maps for two-point targets under varying separations and temporal coding lengths $T_L$, with chirp sequences used as temporal modulation. **e,** SIR of the detection results as a function of $T_L$, comparing two temporal coding schemes: Chirp and binary random sequences. Additional numerical analyses are provided in **Supplementary Note 5**.

**Numerical results.** Numerical experiments were conducted to verify the role of the STC-PM in transforming PR operation into an active-like process. The analysis examines how the coding length $T_L$, the type of temporal modulation (chirp or binary random sequence), the metasurface aperture size, and the observation distance affect object detection performance. As shown in **Fig. 2a**, the STC-PM is positioned at the origin $(0, 0, 0)$ m, and two receivers are placed at $R_1(0, -0.3172, 0.514)$ m and $R_2(0.5, 0, 0)$ m. The metasurface consists of 768 meta-atoms, each with a continuously tunable reflection response over an amplitude-phase range of $[0,1] \times [0°, 180°]$. Three non-cooperative wireless sources are randomly distributed in the surrounding region at $S_1(-0.6, 0, 1.5)$ m, $S_2(0.2, -0.6, 1.7)$ m, and $S_3(0.6, -0.3, 1.2)$ m, producing an interference-rich environment.

To emulate object scanning, a set of STC matrices was predesigned offline and uploaded to the STC-PM, as shown in **Supplementary Video 1**. The metasurface is dynamically reconfigured with a switching interval of 2.5 µs, steering the beam toward one selected source (for example, $S_1$) while treating the others as interference. Each coding matrix generates two distinct spatiotemporal modes: one directs the signal from $S_1$ toward the reference receiver $R_1$, and the other scans the region of interest (**Fig. 2b**). To illustrate the effect of temporal coding, the STC-PM sequentially scans five spatial points (A-E) marked in **Fig. 2a**, with a point-like object located at A. **Fig. 2c** presents the received signals at $R_1$ and $R_2$ and their cross-correlation results $R_{\tilde{y}_1,y_2}(\tau)$. Although the raw echoes at both receivers appear noisy and cluttered, the correlation processing effectively suppresses interference and yields a distinct peak at the target position, confirming that temporal coding enables the separation of object echoes from background clutter and co-channel interference.

The choice of temporal modes $c_{1,2}(t)$ plays a fundamental role in determining the detection performance of an MEPR. To illustrate this effect, we carried out simulations for detecting two point-like objects at varying separations, using chirp sequences as temporal modes with different coding lengths $T_L$. The corresponding detection maps for varying $T_L$ values are shown in **Fig. 2d**, and the dependence of the signal-to-inference ratio (SIR) on $T_L$ is presented in **Fig. 2e**. Both chirp and binary random sequences were evaluated, and additional results are provided in **Supplementary Note 5**, where the influence of observation distance, angular configuration, and metasurface aperture is further analyzed. Three key observations emerge from **Figs. 2c-e**. First, the detection quality improves steadily with increasing $T_L$ and reaches a convergence state beyond a certain coding length. Second, when $T_L$ exceeds approximately $50T_c$, the achievable azimuth resolution of the MEPR becomes comparable to that of an active radar with the same effective aperture, given by $(\lambda/D)R$, where $D$ is the aperture size of the STC-PM, and $\lambda$ and $R$ denote the operating wavelength and the target distance, respectively. Third, when $T_L = T_c$, corresponding to the case of a conventional PR, the detection map exhibits strong clutter and poor discrimination between multiple objects. These findings confirm that the MEPR approaches the detection performance of an active radar once the STC-PM-irrelevant terms $n_{1,2}(t)$ in Eqs. (2)-(3) are effectively suppressed. Increasing the temporal coding length reduces the impact of interference, as evidenced in **Figs. 2d** and **2e**. Overall, the results demonstrate that the spatiotemporal beam manipulation enabled by the STC-PM allows the MEPR to emulate active radar performance while maintaining a passive operational framework.

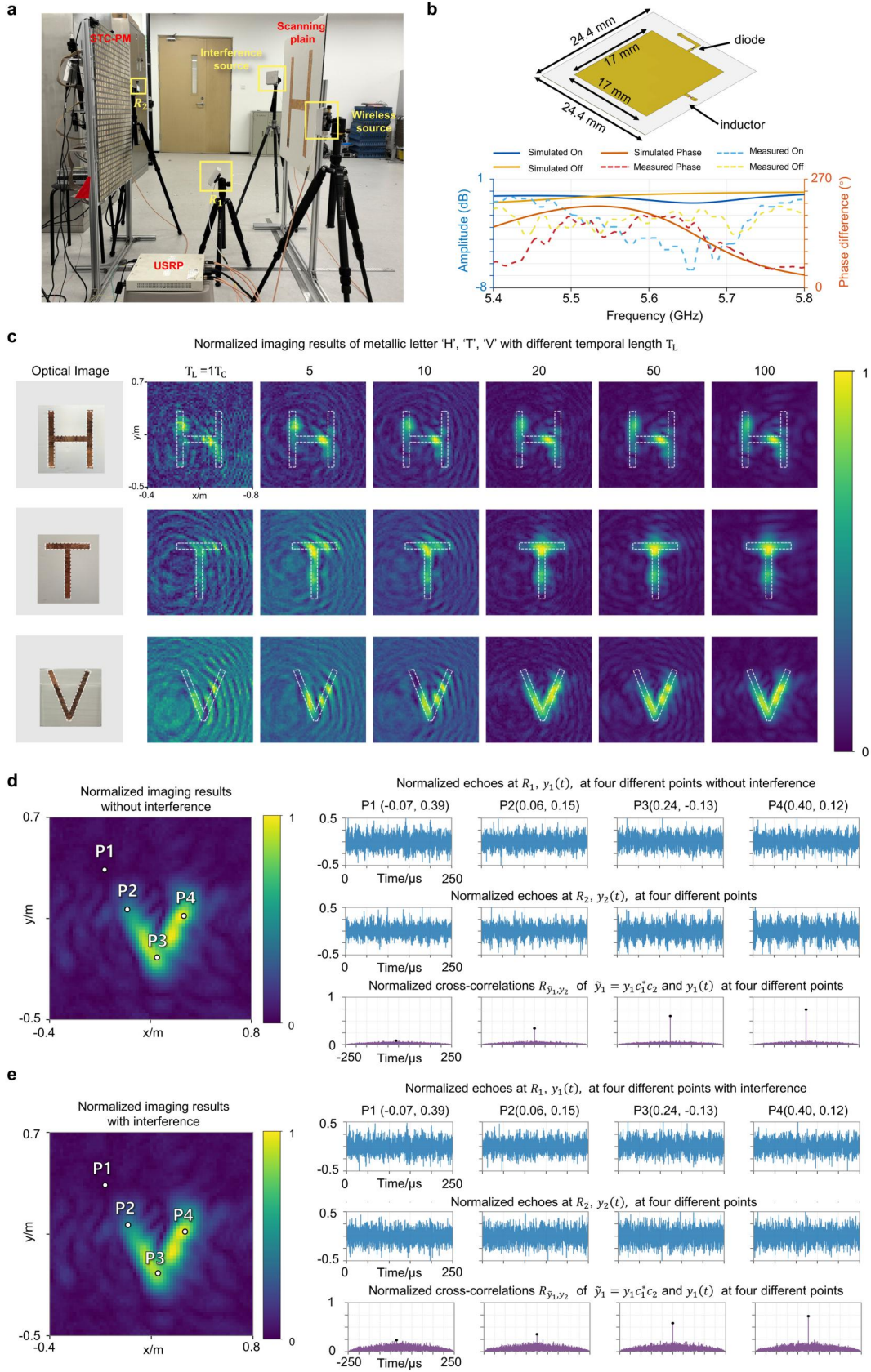

**Figure 3 | Experimental demonstration for near-field imaging.** The experiment employs a home-built one-bit STC-PM consisting of 768 electronically tunable meta-atoms. The metasurface is positioned at the origin (0, 0, 0) m, and two receivers are placed at $R_1(0.5,0,0)$ m and $R_2(0,-0.317,0.514)$ m. Three metallic objects shaped as the letters "H," "T," and "V" are placed 1 m in front of the metasurface within a cluttered indoor environment. The corresponding coding matrices are shown in **Supplementary Video 2**. **a,** Photograph and schematic of the near-field imaging setup. The metasurface dynamically modulates ambient wireless signals to encode spatial information into distinct spatiotemporal signatures, which are captured by the two receivers for correlation-based image reconstruction. **b,** Structure of an individual meta-atom and measured versus simulated magnitude–phase responses across frequency, confirming one-bit phase switching of approximately 180°. **c,** Optical photographs (first column) and reconstructed microwave images (subsequent columns) of the metallic letters obtained under different temporal coding lengths and modulation parameters. Increasing the coding length enhances image contrast and reduces background clutter. **d, e,** Imaging results for the "V" target under interference-free and interference conditions, respectively; the interference is generated by an external source of comparable power to the designated illumination, positioned at $(0.5, 0.3, 1)$ m. Echo signals collected by $R_1$ and $R_2$ at four selected points and their correlation functions show clear peaks at the target location, demonstrating that the MEPR maintains robust imaging performance even in the presence of strong external interference.

**Experimental demonstration of near-field imaging.** To verify that the MEPR can achieve active-like sensing, we experimentally tested its near-field imaging capability using a home-built one-bit STC-PM. The metasurface comprises 768 electronically tunable meta-atoms, each capable of switching between two reflection states corresponding to phase delays of 0° (OFF) and 180° (ON). The structure of an individual meta-atom and its measured and simulated magnitude-phase responses across frequency are shown in **Fig. 3b,** confirming accurate one-bit phase switching of approximately 180° and good agreement between experiments and simulations. This binary control allows the metasurface to dynamically modulate ambient radio signals in both space and time, thereby encoding distinct spatiotemporal signatures into the scattered field. Details of the metasurface design and characterization are provided in the **Methods** section and **Supplementary Note 1**.

As shown in **Fig. 3a**, the experiment was carried out in a realistic laboratory environment containing metal cabinets, chairs, and other reflective objects that introduce strong multipath interference. The STC-PM was positioned at the origin $(0,0,0)$ m, and two receivers were placed at $R_1(0.5,0,0)$ m and $R_2(0,-0.317,0.514)$ m to collect reference and target echoes, respectively. A set of binary-valued STC matrices was predesigned offline to control the metasurface scanning pattern, with a switching interval of 2.5 μs, as shown in **Supplementary Video 2**.

Metallic letters "H," "T," and "V" were used as representative targets to test the imaging capability of the system. The scanning plane, measuring $1.2 \times 1.2\ m^2$ and located 1 m in front of the target, was sampled with a spatial resolution of $0.02\ m \times 0.02\ m$. Each scanning point was measured in 0.25 ms, producing a complete image in 0.93 s.

For the three metallic letter targets ("H," "T," and "V"), the optical photographs are displayed in the first column of **Fig. 3c**, while the corresponding imaging results obtained using the MEPR under different temporal coding lengths are presented in the subsequent columns. Six different temporal coding lengths ($T_L = T_c, 5T_c, 10T_c, 20T_c, 50T_c$, and $100T_c$) were tested. Consistent with theoretical predictions, longer STC sequences lead to clearer object reconstructions and improved contrast. This enhancement arises because longer temporal codes allow the STC-PM to more effectively suppress unwanted components that are unrelated to the metasurface modulation.

Although the indoor environment contains strong background reflections from metallic furniture and surrounding objects, these static clutters can be largely removed through a simple background subtraction procedure similar to that used in conventional active radar processing. To further illustrate the clutter-suppression capability, **Fig. 3d** presents representative echo signals collected by receivers $R_1$ and $R_2$ at four points on the object, along with their cross-correlation results obtained using Eq. (4). While the individual received signals appear noise-like, their correlation yields a distinct peak precisely at the object's location.

Additional experiments were conducted by introducing an external interference source of comparable power to the designated illumination, positioned at $(0.5, 0.3, 1)$ m (see **Fig. 3e**). The MEPR maintained clear detection of the target under these conditions, confirming its robustness against strong co-channel interference. Overall, these results demonstrate that the MEPR accurately reconstructs object profiles and effectively suppress clutter and interference, achieving imaging resolution consistent with theoretical predictions despite operating in a passive configuration.

**Experimental demonstration of far-field UAV tracking.** We experimentally validated the MEPR system for tracking a UAV in flight within a complex indoor environment, as shown in **Fig. 4a**. Details on the UAV platform are provided in **Supplementary Note 6**. The configuration is identical to that in **Fig. 3a**, except that the UAV is positioned in the far field of the STC-PM. From the MEPR's viewpoint, the UAV thus acts as a point-like moving scatterer, analogous to practical low-altitude surveillance scenarios where the target's angular extent is negligible relative to the aperture resolution.

To achieve real-time tracking, a greedy-search algorithm was developed (see **Supplementary Note 7** for details). The system first performs a background scan to record static echoes and construct a reference map for clutter suppression. It then executes a coarse scan, sparsely sampling the field of view to identify potential UAV regions showing enhanced echo responses. Within the most probable region, a fine scan refines the UAV's position by generating a high-resolution local echo map and extracting the energy centroid as the position estimate. Once the UAV is localized by the fine scan, the tracker switches to a greedy, feedback-driven mode that adaptively scans a small local window centered on this position, updating the estimate frame by frame. This adaptive mechanism enables efficient, continuous tracking with high spatial precision. This selective update strategy significantly improves computational efficiency, achieving a tracking update rate of approximately 0.8s per frame.

Representative frames in **Fig. 4c** show the temporal evolution of the UAV trajectory together with synchronized optical recordings. To quantitatively assess the tracking performance, the UAV was programmed to trace trajectories forming the letters "P," "K," "U," "E," "R," and "S" within a horizontal plane at $z = 3.3$m. Under a single-source condition, the reconstructed trajectories closely followed the ground truth (top panels, **Fig. 4d**). When an additional interference source of comparable power was introduced, the MEPR system continued to reconstruct the trajectories accurately (bottom panels, **Fig. 4d**). The error distributions shown in **Fig. 4b** indicate average position errors of 0.116 m without interference and 0.134 m under multi-source interference. Complete tracking sequences are provided in **Supplementary Videos 3–8.**

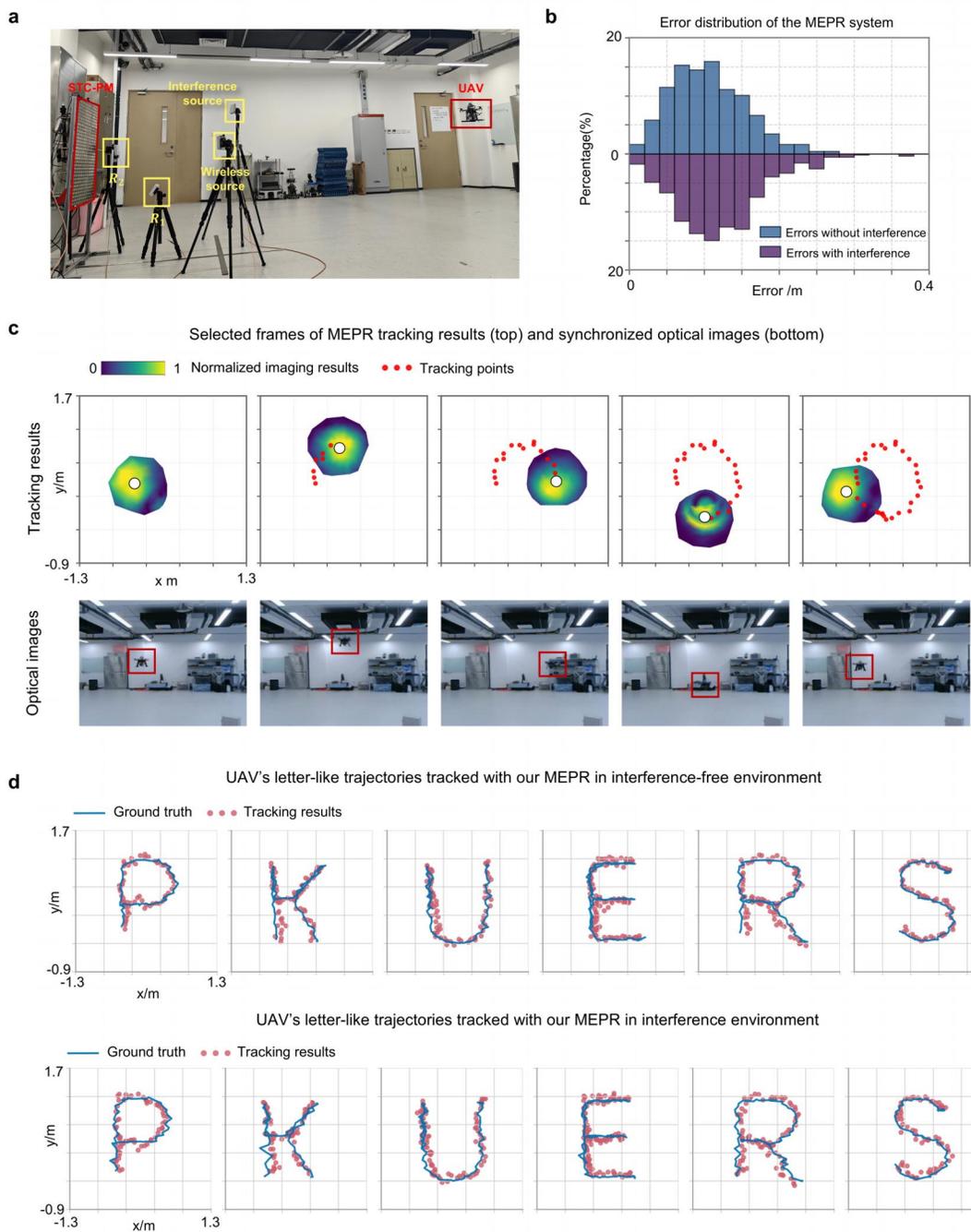

**Figure 4 | Experimental results of far-field UAV tracking**. **a,** Indoor experimental setup for far-field tracking of a UAV flying at a horizontal plane at z = 3.3 m. The MEPR system consists of an STC-PM, a wireless source and two synchronized receivers. The UAV acts as a moving scatterer, while an additional interference source is introduced to evaluate robustness. **b,** Error distributions under interference-free (blue) and interference-present (purple) conditions, illustrating that the system maintains stable localization accuracy even when multiple interference sources are present. **c,** Representative frames of MEPR tracking results (top) and synchronized optical images (bottom), demonstrating continuous trajectory reconstruction of the UAV in real time. **d,** Tracking results for

UAV trajectories forming the letters "P," "K," "U," "E," "R," and "S." The top row shows results without interference, and the bottom row shows results with multiple interference sources. The MEPR system accurately reconstructs the UAV trajectories under both conditions.

Overall, these results highlight how spatiotemporal modulation enables the MEPR to isolate and follow weak moving targets using only ambient radio signals. By embedding programmable coding into the propagation channel, the MEPR effectively endows the environment itself with sensing capability, pointing toward a new paradigm of intelligent and interference-resilient EM perception without active transmission.

## Discussion and conclusions

The proposed MEPR introduces a new approach to passive sensing by integrating PMs capable of space-time modulation. In contrast to conventional PRs that depend on Doppler analysis and demodulation of non-cooperative signals, the MEPR embeds active-like modulation directly into target echoes, enabling their separation from strong interference such as direct-path and co-channel signals. Experimental results confirm the feasibility and robustness of this concept, achieving sub-meter localization accuracy and real-time UAV tracking in interference-rich environments.

Although these results are promising, several challenges remain. The present experiments were conducted indoor under controlled conditions, while outdoor or mobile operation may introduce additional factors such as time-varying multipath and platform instability. Moreover, the current implementation requires a pre-acquired background map to remove static clutter, which may limit adaptability in dynamic scenes. Future research will address these challenges through adaptive background modeling, distributed and large-aperture metasurface architectures, and multi-target tracking strategies.

Overall, the MEPR establishes a foundation for high-performance, low-profile, and scalable PR systems. By enabling programmable control of the spatiotemporal characteristics of ambient EM fields, it bridges the conceptual gap between passive and active detection. This approach opens opportunities for intelligent EM environments capable of perception, adaptation, and interaction, with broad implications for UAV surveillance, border security, and next-generation situational awareness technologies.

## Methods

**One-bit STC-PM.** The STC-PM operates at 5.48 GHz, within a frequency band widely used for wireless communication and radar sensing. It consists of a $32 \times 24$ array of PIN-diode-loaded meta-atoms, forming an effective aperture of $0.781 \times 0.586$ m². Each meta-atom contains a PIN diode (SMP1345-079LF) connecting a metallic resonant patch and a microstrip phase-shifting line. By switching the bias voltage between 12 V (ON) and 0 V (OFF), the metasurface generates two discrete reflection phases (0° and 180°), enabling one-bit programmable control of the reflected wavefront.

The array is driven in real time by a field-programmable gate array (FPGA) linked to a host computer, supporting sub-microsecond switching for static and dynamic space–time modulation. Each unit cell uses a dual-layer substrate, with F4B $\varepsilon_r = 2.55$, $\tan\delta = 0.0019$) providing low-loss EM performance and FR4 providing mechanical stability, and includes a 33 nH RF choke for bias isolation. The electromagnetic response was verified by full-wave simulations (CST Studio Suite 2017) and confirmed by free-space measurements, both showing a stable 0°/180° phase contrast across the operational band. Further design and implementation details are provided in **Supplementary Note 1**.

**$K$-component spatiotemporal modal decomposition.** To describe the operation of the STC-PM, we introduce a $K$-component spatiotemporal modal decomposition model. The STC-PM is treated as a 2D array of $N$ controllable meta-atoms, each characterized by a time-varying EM response $\Gamma(n,t)$, where $n$ denotes the index of the meta-atom. The response can be expressed as

$$\Gamma(n;t) = \sum_k^K f_k(n) c_k(t) \tag{5}$$

where $f_k(n)$ and $c_k(t)$ represent the $k$-th spatial and temporal modes, respectively, and $K$ is the total number of spatiotemporal modes.

When the STC-PM is illuminated by a radio signal $(t; \boldsymbol{r}_s)$ from a wireless source located at $\boldsymbol{r}_s$, the response at an observation point $\boldsymbol{r}$ can be written as

$$y(t;\boldsymbol{r},\boldsymbol{r}_s) = \sum_{k=1}^{K}\left[\underbrace{\int d\omega \tilde{s}(\omega)e^{j\omega t}\overbrace{\sum_{n=1}^{N}f_k(n)\widetilde{H}_{r_s\to n\to r}(\omega)}^{\tilde{F}_k(\omega;\boldsymbol{r},\boldsymbol{r}_s)}}_{s_k^h(t;\boldsymbol{r},\boldsymbol{r}_s)=f_k(t;\boldsymbol{r},\boldsymbol{r}_s)\otimes s(t;\boldsymbol{r}_s)}\right]c_k(t)$$

$$= \sum_{k=1}^{K} s_k^h(t;\boldsymbol{r},\boldsymbol{r}_s)\, c_k(t) \tag{6}$$

Equation (6) assumes a single wireless source, but can be extended to multiple sources through linear superposition, $s_k^h(t;\boldsymbol{r},\boldsymbol{r}_s) \to \sum_s s_k^h(t;\boldsymbol{r},\boldsymbol{r}_s)$. In this formulation, the time origin is taken at the metasurface. If referenced to the observation point $\boldsymbol{r}$, the temporal modes are modified as $c_k(t) \to c_k(t - \tau_{m\to r})$, where $\tau_{m\to r}$ denotes the propagation delay from the STC-PM to $\boldsymbol{r}$.

Two remarks can be made from Eq. (6). First, the *K*-component decomposition provides a general theoretical framework for describing spatiotemporally modulated metasurfaces. Specific implementations, such as frequency-scanning antennas, can be obtained by setting $c_k(t) = e^{j\omega_k t}$ with $\omega_i \neq \omega_j$ for $i \neq j$, and choosing $f_k(n)$ to form independent radiation beams, $\sum_{n=1}^{N} f_k(n) \widetilde{H}_{r_s\to n\to r}(\omega)$. Second, in free space, the channel response can be approximated as $h_k(t;\boldsymbol{r},\boldsymbol{r}_s) \approx G_k \delta(t - \tau_{m\to r})$, where $G_k$ denotes the spatial radiation gain of the *k*-th mode. Substituting this into Eq. (6) yields the signal models described by Eqs. (2) and (3), which account for additional non-metasurface-related interference terms $n_1(t)$ and $n_2(t)$. Further details are provided in **Supplementary Notes 2 and 3**.

## Data Availability:

The data that support the findings of this study are available from L. L. upon request.

## Code Availability:

Code that supports the findings of this study is available upon reasonable request from L. L.

## Acknowledgements:

This work was supported by the National Key Research and Development Program of China under Grant Nos. 2021YFA1401002, 2023YFB3811502. T. J. C. acknowledges the support from the National Natural Science Foundation of China under Grant No. 62288101. V. G. acknowledges partial support from the European Union–Next Generation EU under the Italian National Recovery and Resilience Plan (NRRP), Mission 4, Component 2, Investment 1.3, CUP E63C22002040007, partnership on "Telecommunications of the Future" (PE00000001 - program "RESTART"), and from the University of Sannio under the FRA program.## Author Contributions Statement:

L. L., T. J. C. and V. G. conceived the idea, and wrote the manuscript. M. L., J. X. H. Z. and X. Z. designed and developed the system and conducted the experiments. All authors participated in the data analysis and interpretation, and read the manuscript.

## Completing Interests Statement:

The authors declare no conflicts of interests.